\documentstyle[11pt]{article}
\begin{document}

\title{Optimizing Conflicts\\
in the Formation of Strategic Alliances }

\author{
Razvan Florian$^{1}$\footnote{florian@arxia.com} and Serge
Galam$^{2}$\footnote{galam@ccr.jussieu.fr} \\ \\ $^1$\'Ecole
Normale Sup\'erieure de Cachan, D\'epartement de Physique,\\ 61,
avenue du Pr\'esident Wilson, 94235 Cachan, France\\ \\
$^2$LMDH\footnote{Laboratoire associ\'e au CNRS (UMR n$^{\circ}$
800)}, Universit\'e Paris 6, Case 86, \\ 4 place Jussieu,75252
Paris cedex 05, France }
\date{February 10, 2000}

\maketitle

\begin{abstract}
Coalition setting among a set of actors (countries, firms,
individuals) is studied using concepts from the theory of spin
glasses. Given the distribution of respective bilateral
propensities to either cooperation or conflict, the phenomenon of
local aggregation is modeled. In particular the number of
coalitions is determined according to a minimum conflict
principle. It is found not to be always two. Along these lines,
previous studies are revisited and are found not to be consistent
with their own principles. The model is then used to describe the
fragmentation of former Yugoslavia. Results are compared to the
actual situation.


\end{abstract}

\section{Introduction}

Physical concepts might prove useful in describing collective
social phenomena. Indeed models inspired by statistical physics
are now appearing in scientific literature \cite{Stauffer}.

The process of aggregation among a set of actors seems to be a
good candidate for a statistical physics like model. These actors
might be countries which ally into international coalitions,
companies which adopt common standards, parties that make
alliances, individuals which form different interest groups, and
so on.

Given a set of actors, there always exists an associated
distribution of bilateral propensities towards either cooperation
or conflict. The question then arises as to how to satisfy such
opposing constraints simultaneously. In other words, what kind of
alliances, if any, will optimize all actor bilateral trends to
respectively conflict or cooperation.

It turns out that a similar problem does exist in spin glasses.
For these systems, magnetic exchanges are distributed randomly
between ferro and antiferromagnetic couplings. Indeed such an
analogy has been used in the past in a few models.

The first one is by Axelrod and Bennett (hereafter denoted as AB)
\cite{Landscape}. They apply the physical concept of minimum
energy to single out the stable coalitions within an configuration
landscape. Later Galam demonstrated that the AB model is not fully
consistent with its physical content. He then suggested a second
model (hereafter denoted as G) based on both random bond and
random site exchanges \cite{Fragmentation}.

We propose here a qualitative extension of defining the coalition
problem. While all previous approaches used implicitly a 2-side
coalition dynamics, we allow multi-side coalitions. We indeed
define the actual number of alliances as an internal parameter of
the problem to be determined by simultaneously optimizing all
bilateral propensities. While an Ising like variable may be
appropriate to a huge number of problems both in physics and
outside physics, there is no a priori reason to consider coalition
dynamics restricted to a bimodal distribution.

We then apply our model to study the fragmentation of former
Yugoslavia. Results are compared to the actual situation. We also
revisit previous studies along this new scheme of not setting a
priori the number of coalitions. As soon as the artificial bimodal
constraint is relaxed, their results are no more consistent with
reality.

Moreover, in physics models the exchange couplings and their
distributions are known. In contrast, in the coalition problem the
quantitative determination of bilateral propensities to either
cooperation or conflict is a major challenge. We discuss this
matter and argue that previous propensity calculations were too
simplistic.

The rest of the paper is organized as follows. In the second part
we review the AB and G models. The qualitative extension to
multimodal  coalitions is discussed in Section 3. On this basis,
we revisit in Section 4 the cases of the Second World War and the
Unix standard setting (studied in Ref. \cite {Landscape} and
respectively \cite{Unix}). The case of the fragmentation of former
Yugoslavia is analyzed in Section 5. Concluding remarks are
contained in last Section.

\section{Former models}

\subsection{The Axelrod-Bennett model (AB)}

Axelrod and Bennett first attempted \cite{Landscape} to explain
the composition of coalitions using some characteristics of the
involved actors (the elements of the system). The relative
affinity between actor $i$ and actor $j$ is measured by a pairwise
propensity $p_{ij}$, which is negative when they are in conflict
and positive when they like to cooperate (with $p_{ii}=0$).
Differences in actor sizes are expressed by assigning to each one
a weight factor $s_{i}$ (a positive quantity). It may be a
demographic, economic or military factor, or an aggregate
parameter.

The creation of coalitions among the actors introduces a distance
$d_{ij}$ between each pair $(i,\ j)$ of actors. It is 0 if $i$ and
$j$ belong to the same coalition and 1 when they are in different
coalitions. A given partition $X$ of the actors into coalitions is
equivalent to knowing all $d_{ij}$.

A ``frustration" of each actor $i$ is introduced to measure how
much a configuration $X$ satisfies its propensities. It writes,
\begin{equation}
F_{i}=\sum _{j=1}^{n}s_{j}p_{ij}d_{ij}(X)\ ,
\end{equation}
where we considered a group of $n$ actors.

Adding all actor frustrations, respectively weighted by their
sizes, results in an ``energy" of the system,
\begin{equation}
E(X)=\frac{1}{2}\sum _{i=1}^{n}s_{i}F_{i}\ ,
\end{equation}
which is,
\begin{equation}
\label{eqenergy} E(X)= \sum _{i>j}^{n} s_{i}s_{j}p_{ij}d_{ij}(X)\
.
\end{equation}

It is then postulated that the actual configuration of the system
is the one which minimizes the energy. The path followed by the
system into the coalition landscape space from an initial
configuration follows the direction of the greatest gradient of
energy. Once a minimum is reached the system does not change.

The AB model has been applied to the study of the alliances of the
Second World War \cite{Landscape}. It was also used to study the
standard setting coalitions formed by the companies which
developed the Unix operating system \cite{Unix}.

Later Galam has demonstrated several inconsistencies of the AB
model \cite{Comment}. Here we found additional setbacks (see
Section 4).

\subsection{Galam reformulation of the AB model}

Galam has shown \cite{Fragmentation} that in case of bimodal
 coalitions ($A$ and $B$),
the AB model is totally equivalent to a finite spin glass at zero
temperature.

Accordingly, configurations can be expressed by the spin variables
$\eta _{i}$, where the spin is +1 if the actor $i$ belongs to
coalition $A$ and it is -1 if the actor belongs to $B$. The
distances can be rewritten as $d_{ij}=\frac{1}{2}(1-\eta _{i}\eta
_{j})$. The energy becomes,
\begin{equation}
E(X) = E_{0} - \frac{1}{2} \sum _{i>j}^{n} J_{ij} \eta _{i} (X)
\eta _{j} (X) \ ,
\end{equation}
where $J_{ij} = s_{i} s_{j} p_{ij}$ and
\begin{equation}
\label{e0}
 E_{0} = \frac{1}{2} \sum _{i>j}^{n} J_{ij}
\end{equation}
is a constant which depends only on the given sizes and
propensities, but not on the configurations.

\subsection{The Galam model (G)}

Galam \cite{Fragmentation} introduced a new model for the case of
bimodal coalitions where the two alliances have clear
characteristics that would allow to define for each actor a
natural belonging $\epsilon _{i}$ to each alliance. It is +1 if
the actor $i$ should be in coalition $A$, according to its own
characteristics compared with those of $A$. It is -1 for $B$ and 0
if there is no natural belonging to neither $A$ or $B$.

In the G model, the benefit $J_{ij}$ of cooperation or conflict
between a pair of countries in either the same or opposite
alliance is accounted for in addition to the local bond
propensities $G_{ij}$. Both add together to a total propensity
$p_{ij}=G_{ij}+\eta _{i} \eta _{j} J_{ij}$.

In parallel, forces (like military or economic mechanisms) by
which each coalition as a whole couples to the orientation of a
given actor are expressed in terms of an external field $\beta
_{i}$. It contributes to the overall conflict through the product
$\beta _{i} b_{i}\eta _{i}$, where $\beta _{i}=\pm 1$ represents
the direction of the force that acts on actor $i$ (towards $A$ or
$B$), while $b_{i}>0$ is the amplitude of this force. The total
energy of the system sums up to
\begin{equation}
E=-\frac {1}{2}\sum _{i>j}^{n}(G_{ij} +\epsilon _{i}\epsilon
_{j}J_{ij})\eta _{i}\eta _{j} -\sum _{i}^{n}\beta _{i} b_{i}\eta
_{i} \ .
\end{equation}

This model has been used \cite{Fragmentation} to explain the
stability of alliances during the cold war, with opposing NATO and
Warsaw pacts. The fragmentation of Eastern Europe which resulted
from the Warsaw pact dissolution as well as the simultaneous
western stability was recovered within the model. The European
construction versus stability and the Chinese stability were also
analyzed within the same model \cite{Fragmentation}.

\section {Beyond previous models}

\subsection{A new approach: going multimodal}

In physics, the symmetry of the internal degrees of freedom of a
given system is clearly determined from specific measurements on
the studied material. In particular, most magnetic systems may be
modeled using a spin variable. It can be, among others, a 2-state
Ising variable, a q-state Potts model, a XY spin with planar
symmetry, or a Heisenberg spin with continuous symmetry.

In the case of human systems, the a priori restriction of using an
Ising variable has no ground. Some known coalitions exhibit more
than 2 simultaneous alliances, though not much more. For example,
the users and the producers of personal computers can be divided
in at least three categories, with regard to the operating system
they use (Windows, Mac OS, Linux $\backslash$ Unix); there were
three groups that fought each other between 1941 and 1945 in the
space of ex-Yugoslavia: the Chetnik guerilla groups, the Communist
bands and the Nazis \cite{Yug2}; and in most democratic countries
there are more than two independent political parties.

We extend previous approaches by allowing for multimodal
coalitions. We assume the number $q$ of coalitions to be an
internal degree of freedom to vary from 1 up to $n$. Before, $q$
was arbitrarily fixed to 2. The spin variable $\eta _{i}$ can thus
take $q$ different values. Distances are expressed as
$d_{ij}=1-\delta (\eta _{i}, \eta _{j})$ and the energy is
\begin{equation}
\label{eqmult}
 E(X)=2 E_{0}-\sum _{i>j}^{n}J_{ij} \delta (\eta
_{i}(X), \eta _{j}(X)) \ ,
\end{equation}
where $E_{0}$ is given by equation (\ref{e0}). This corresponds to
the $q$ state Potts model \cite{Potts}. The initial expression of
the energy, equation (\ref{eqenergy}), can still be used.

To demonstrate that the increase in the number of allowed
coalitions may reduce the energy of the system, we will discuss
the example of a system with $n=3$ actors. We assume $J_{1 2}=J_{2
3}=J_{1 3}=-1$.

If we impose only one coalition, the energy is 0. If we impose two
coalitions, the minimum energy is -2 and is three times
degenerated (the system is frustrated). If we don't impose a fixed
number of coalitions, the system sets itself in a configuration
with three different coalitions (all the actors are independent),
with an energy -3, the lowest energy possible.

\subsection{Gauge transformations}
Let us consider a transformation of the distances as $d'_{ij}=a \,
d_{ij} + b$. The energy becomes
\begin{equation}
E'(X)= \sum _{i>j}^{n} J_{ij}d'_{ij}(X) =a \, E(X) + 2b \, E_{0} \
.
\end{equation}
If $a>0$ this transformation keeps unchanged the dynamics of the
system.

An interesting transformation is given by $d'_{ij}=2 d_{ij} -1$.
That is $d'_{ij}=-1$ if $i$ and $j$ are in the same coalition and
$d'_{ij}=1$ if not. For bimodal coalitions, we have $d'_{ij}=-\eta
_{i} \eta _{j}$. The energy expressed with this distance accounts
equally for conflict and cooperation.

Another interesting transformation is $d''_{ij}=d_{ij} -1$, which
gives $d''_{ij}=-\delta(\eta _{i}, \eta _{j})$. In this case the
energy is exactly the Potts energy,
\begin{equation}
E''(X)= \sum _{i>j}^{n} J_{ij}d''_{ij}(X) = -\sum _{i>j}^{n}J_{ij}
\delta (\eta _{i}(X), \eta _{j}(X)) \ .
\end{equation}
The minimization of this energy maximizes the cooperation.

Because all these distances yield the same dynamics of the system,
it doesn't need to be specified exactly if the actor aim is,
psychologically, the minimization of conflicts, the maximization
of cooperation, or both, and in which proportion. However, this
prevents us from comparing the energies of different systems
(including the same actors with some propensities changed) until
this is established. In our study we will stick to the original AB
form of the distance (which minimizes conflicts).

\subsection{Neutrality}

The introduction of spin variables allows us to consider the
possibility of neutrality by letting the spins to have also the
value $\eta _{i}=0$. In the case of bimodal coalitions, this is
straightforward. In the case of $n$ coalitions, the energy
(\ref{eqmult})  has to be rewritten as
\begin{equation}
E=2 E_{0}-\sum _{i>j}^{n}J_{ij} \delta (\eta _{i}, \eta
_{j})[1-\delta(\eta _{i}\eta _{j},0)] \ ,
\end{equation}
where the last factor accounts for the case when both $i$ and $j$
are neutral.

\section {Application to real cases}

\subsection {The case of the Second World War}

Axelrod and Bennett have applied their model of aggregation to
explain the composition of the opposing alliances during the
Second World War \cite{Landscape}. The actors are 17 European
countries involved in the war. Country sizes are measured with a
national capabilities index which combines six components of
demographic, industrial and military power. Propensities are
computed  from 1936 data. The criteria used are: ethnic conflicts,
religion, border disagreements, type of government and history of
war. They are combined with equal weights.

After numerical minimization of the energy, two different minima
were found. The absolute one corresponds to respectively Britain,
France,  Soviet Union, Czechoslovakia, Yugoslavia, Greece and
Denmark in one coalition versus Germany, Italy, Poland, Romania,
Hungary, Portugal, Finland, Latvia, Lithuania and Estonia in the
other one. After the criteria of Axelrod and Bennett (who measure
the alignment by whether a country was invaded by another country,
or had war declared against it), this corresponds to the
historical reality, with the exception of Poland, which is on the
wrong side, and of Portugal, which was neutral. There exists
another local minimum with Soviet Union, Yugoslavia and Greece
versus all the others.

While Axelrod and Bennett exhibit this result as a validation of
their model, they hardly comment a crucial assumption they made.
They supposed that the 17 countries can be partitioned in only 2
coalitions. Even if this was the historical reality, this
artificially added constraint does not fit to the principles of
the model. The final configuration should be determined only by
the actors' pairwise interactions, via the minimization of the
energy. Nothing restricts a priori the existence of more than 2
coalitions. A partition in two coalitions should be indeed a
result of the model and a proof of its predictiveness.

\subsection {Revisiting the Second World War case}

We have redone the AB computation using the same data set for
sizes and propensities (which is available on Internet
\cite{web}).

We first confirmed the AB results once the number of coalitions is
restricted to 2. We then introduced the possibility of spin 0 in
the hope of capturing the neutrality of Portugal while keeping the
other conditions unchanged. The results stay unchanged with no
country neutral.

We then allowed an a priori unlimited number of coalitions (from 1
up to $n=17$). Exploring then the full energy landscape, we found
that there exists one single minimum associated to 3 alliances.
They are respectively Soviet Union and Greece versus Germany,
Italy, Estonia and Latvia versus Britain, France, Czechoslovakia,
Denmark, Yugoslavia, Poland, Romania, Hungary, Portugal, Finland
and Lithuania. This minimum has an energy $E=-132.36$. It is  much
less than the energies of the two AB configurations which are
respectively -94.23 and -91.06.

We also tried to find a 2 coalition configuration imposing
geographical constraints, but the result stays a three alliance
configuration. This shows that, in fact, the real prediction of
the AB model, using the same propensities, is rather far from the
historical reality.

\subsection {The Unix case}

Axelrod et al. applied the AB model of aggregation to explain also
the formation of standard setting alliances, like for the Unix
operating system \cite{Unix}. The actors are 9 companies involved
in the development of Unix (AT\&T, Sun, Apollo, DEC, HP,
Intergraph, SGI, IBM and Prime). Sizes are given by the firms'
share in the technical workstation market or by expert estimates
(for AT\&T).

If a firm $i$ belongs to alliance $L$, its utility (the
satisfaction of the economic agent, given by its profit) is
expressed by
\begin{equation}
U_{i}(L)=\sum _{j\in L} s_{j}-\left[ \alpha \sum _{j\in L'}
s_{j}+(\alpha + \beta) \sum _{j\in L''} s_{j} \right] \ ,
\end{equation}
where $s_{j}$ is the size of firm $j$, and $L'$ and $L''$ form a
partition of alliance $L$ into distant and close rivals of $i$
($L=L'\cup L''$ and $L'\cap L''=\emptyset$). The parameters
$\alpha$ and $\beta$ are positive. The identification of close and
distant rivals is done from the degree of specialization of the
firms in the production of Unix-based workstations.

We may rewrite the utility as
\begin{equation}
U_{i}(L)=\sum _{j\in L} s_{j}p_{ij}=\sum _{j=1}^{n} (1-d_{ij})
s_{j} p_{ij} \ ,
\end{equation}
where the last expression gives the utility of $i$ without having
to specify its alliance. This leads to the propensities
$p_{ij}=1-\alpha$, if $i$ and $j$ are distant rivals, and
$p_{ij}=1-(\alpha+\beta)$, if $i$ and $j$ are close rivals.

The energy (\ref{eqenergy}) equals
\begin{equation}
E=2 E_{0}-\frac{1}{2}\sum_{i=1}^{n}s_{i}U_{i} \ ,
\end{equation}
so a minimization of the energy yields a weighted maximization of
the utilities.

For the choice of parameters $\alpha=\beta=1$ (and in all cases
for which $0.8\leq\alpha\leq 1.5$, $0.7\leq\beta\leq 1.5$),
Axelrod et al. found 2 minimums of the energy with the same value.
One configuration is Sun, DEC and HP versus AT\&T, Apollo,
Intergraph, SGI, Prime and IBM. The other one is Sun, AT\&T, Prime
and IBM versus DEC, HP, Apollo, Intergraph and SGI. The latter
configuration corresponds to Unix International vs. Open Software
Foundation, and only IBM is incorrectly assigned.

Axelrod et al. advocate this result in favor of the effectiveness
of their methodology. However, they have again imposed the
artificial constraint of allowing only a maximum of 2 coalitions.
They motivate this choice by the fact that ``the positive
externality created by standardization declines as the number of
standards increases and thereby reduces the principal advantage of
setting standards, which is a larger post-standardization market"
(\cite{Unix}, p. 1498).

But this condition is in fact included in the formula of the
utility $U_{i}(L)$, which grows linearly with the total size of
coalition $L$. Therefore, the aggregation of firms into a small
number of alliances should be only the result of the energy
minimization. If the utility grows faster than linearly with the
coalition size (which may be the case in reality), and the linear
approximation is not good enough, then even if we impose a maximum
of 2 coalitions the result would be wrong, because of the
discrepancy between the real and the modeled utility.

\subsection {Revisiting the Unix case}

We checked the above model with no conditions imposed on the
coalition number. For the case $\alpha=\beta=1$ there exist 150
different minimums of the energy, with an average of 6.57
coalitions per configuration. This is definitely a bad choice of
parameters because all the propensities are negative or 0, so
there is little incentive for aggregation.

The condition for the beginning of aggregation is $\alpha<1$. We
sampled the interval $0\leq\alpha\leq 0.9$, $0\leq\beta\leq 2$,
with a step of 0.1. For $\alpha+\beta<1$, all firms form a single
block (all propensities are positive).  Configurations with 2
coalitions appear in an interval given roughly by
$1-\alpha<\beta<2(1-\alpha)$.

There are 14 types of configurations of 2 coalitions. We ranked
their importance from the number of pairs ($\alpha$, $\beta$) for
which they are realized, weighted with the size of their basin of
attraction. The most preferred configurations are (i) Sun, AT\&T
and IBM versus  the others (29.6 \%), (ii) AT\&T and Apollo versus
the others (28.4 \%), (iii) IBM, DEC and Apollo versus the others
(12.5\%), the rest having less than 10 \% each.

The ``best" configuration predicted by Axelrod et al. is not
realized at all, while the other configuration they predicted has
rank 4 (5.5\%). The real empirical configuration Sun, AT\&T and
Prime versus the others has rank 9 (2.7\%) and doesn't have a
basin of attraction greater than 6\% for any combination of
parameters.

Again, as for the Second World War, the prediction of the non
constrained AB model, using the same propensities, doesn't fit the
empirical reality.

\section {A new application: the case of Yugoslavia}

We have studied a new problem, the fragmentation of the former
Yugoslavia. The actors are its 8 administrative entities
(provinces or republics): Serbia, Croatia, Bosnia, Slovenia,
Macedonia, Vojvodina, Kosovo and Montenegro. Here a coalition
means a federation of entities, or an independent state if it has
only one member.

In the initial configuration all the actors are in the same
coalition - the former Yugoslavian federation. The system is then
left to evolve in the direction of the greatest gradient of energy
down to the minimum which should correspond to the stable
configuration.

To implement our model we needed first to evaluate all the
propensities among the set of the 8 entities. The ethnic group
diversity \cite{Yug2} of the whole set is a major ingredient which
we considered in evaluating those propensities. We also took into
account differences in religion and language. Entity sizes are
taken proportional to population sizes.

We used the 1981 census results \cite{Yug1}, \cite{Yug2}. We
considered eight major ethnic groups: Serbs, Croats, Muslims,
Slovenes, Macedonians, Montenegrins, Albanians and Hungarians. We
neglected in our study the influence of other ethnic groups. They
accounted for less than 1\% each of the total population of
Yugoslavia in 1981. We also neglected the 5.4\% of the population
which label themselves Yugoslavian. We assumed this later group to
be perfectly tolerant of the others (zero propensity).

Due to lack of accurate data for all the entities, we considered
all the Serbs, Macedonians and Montenegrins to be Orthodox
Christians; all the Croats, Slovenes and Hungarian to be Catholic
Christians; and all the Muslims and Albanians to be Muslim. For
1999, there exist differences up to 25\% between the figures for
nationality and religion, but in most cases they are much less
\cite{CIA}. We classified the ethnic groups, with regard to their
language, as Serbo-Croats (Serbs, Croats, Muslims and
Montenegrins), other Slavs (Macedonians and Slovenes) and non
Slavs (Albanians and Hungarians) \cite{Yug2}.

The propensities between pairs of entities are computed as
follows:
\begin{equation}
p_{ij}=\sum _{k, l}^8 q_{ik} q_{jl} w_{kl} \ ,
\end{equation}
where $q_{ik}$ represents the percentage of ethnic group $k$ in
entity $i$ and $w_{kl}$ represents the pairwise propensity between
ethnic groups $k$ and $l$.

For $k=l$, $w_{kk}=+1$. For $k\neq l$, the $w_{kl}$'s are computed
as the sum of 2 terms. One stands for religion and the other for
language: $w_{kl}=\omega _{religion}(k, l)+\omega _{language} (k,
l)$.

We used the hypothesis that Christians are more tolerant of other
Christian religion members than for Muslims. We also assumed that
the Serbo-Croats are more tolerant of other Slavs than of non
Slavs.

For religion, the factor $\omega _{religion}$ is positive and
equals $+\omega _{1}$ for pairs of ethnic group with the same
religion. It is negative and equals $-\omega _{2}$ for pairs of
Catholic Christian and Orthodox Christian ethnic groups. For pairs
of Christian and Muslim groups, the factor is $-\omega _{3}$.

For language, the factor $\omega _{language}$ is $+\omega _{4}$ in
the case of two Serbo-Croatian speaking groups. For pairs of two
different Slav groups, the factor is $-\omega _{5}$. For pairs
that include at least a non Slavic language, the factor is
$-\omega _{6}$. All the $\omega _{i}$'s are positive.

The graduation of tolerance previously described yields the
following conditions: $\omega _{2}<\omega _{3}$ and $\omega
_{5}<\omega _{6}$. We also have the condition $\omega _{1}+\omega
_{4}<1$ to prevent a pairwise propensity between 2 different
ethnic groups to be greater than the propensity within the same
group, which is the reference factor for other propensities.

The parameters $\omega _{i}$ are unknown. We varied those
parameters in the domain $0<\omega _{1,4}<0.5$; $0<\omega
_{3,6}<1$; $\omega _{2}<\omega _{3}$; $\omega _{5}<\omega _{6}$,
with a step of 0.05 and checked the results predicted by the
minimization of the energy. There are 28 resulting configurations
(from a total of 4140 possibilities).

All these configurations respect the geographical connectivity. We
ranked them from the number of cases they are realized, weighted
with the relative size of their basin of attraction in each case.
Only three configurations appear in more than 10\% of cases each
and 11 other appear in more than 1\% of cases each.

The main three configurations are: (i) A federation including
Serbia, Croatia, Bosnia, Montenegro and Vojvodina, with the other
entities being independent, is obtained in 42.4\% of cases. (ii)
In 12.1\% of cases the result is Serbia, Montenegro and Vojvodina,
and the others independent. (iii) In 10.9\% of cases, only Serbia
and Vojvodina stay together.

The above results yield the real configuration which resulted from
the fragmentation of Yugoslavia for the second largest set of
parameters. We considered Kosovo practically separated from Serbia
because of its special situation after the 1999 war, under the
control of NATO peacekeepers.

For this final configuration, in 31\% of cases (for example for a
choice of parameters $\omega _{i}$ like (0.2, 0.5, 0.6, 0.1, 0.3,
0.5)), the order of fragmentation is the real one: Croatia and
Slovenia - June 1991, Macedonia - September 1991, Bosnia - April
1992, with the general exception of Kosovo. Our model predicts its
splitting from Yugoslavia on the first or the third step, due to
its mostly Albanian population. Within our model and the G model
we might say that Kosovo was artificially kept inside Serbia using
an external field. The field was then destroyed by the NATO
bombing.

\section {Conclusion}

The introduction of analogies inspired from spin glass models into
the study of social systems makes possible the prediction of the
dynamics of macroscopic alliances formation within a given system
of actors. While pairwise interactions are clearly instrumental in
this approach, there is no scientific method to date to select the
factors to be accounted for in their evaluation.

We have shown in particular that previous studies were not
consistent with their own principles. Using the propensities and
sizes computed from their methodology and then computing the
resulting configurations, following strictly and solely the
principle of the minimum energy, we have shown that the results
don't fit any more to reality. Nevertheless we have also shown
that there exist some parameter ranges which do yield the real
situation.

We have also emphasized that analogies with physics should not
mean a straight mapping. A major difference here is to consider
the number of alliances to be a free internal parameter of the
system. In physics it is predetermined.

In conclusion our approach validates the feasibility of modeling
strategic international behavior, but at the same time it
demonstrates also the dangers of taking it too simplistically.
Within our model, we have also reported configurations which don't
fit to reality. At this stage, this shows that more work is
necessary to single out a feasible scheme for the evaluation of
parameters.

\end{document}